\documentclass{eptcs}
\providecommand{\event}{SCSS 2012} 
\usepackage{breakurl}             

 \usepackage{multicol}
\usepackage{xkeyval}

 \usepackage{amsmath}
\usepackage{longtable}
 \usepackage[utf8x]{inputenc} \usepackage[T1]{fontenc}
\usepackage{coqdoc}
\usepackage[all]{xy}
\usepackage{paralist}
\usepackage{pdflscape}
\usepackage{tikz}
\usetikzlibrary{decorations.markings}

\usetikzlibrary{positioning}

\newcommand{\nouvelleN}[3]{
\tikzstyle{noeud}=[ inner sep=1pt, outer sep= 1 pt,
       rectangle,rounded corners=10pt,draw,
       fill=pink!75,text=red]
\node[noeud] (#2) [#1] {\begin{minipage}{.25\textwidth} #3
 \end{minipage}};
}

\usetikzlibrary{arrows}

\newtheorem{definition}{Definition}

\title{Performing Implicit Induction Reasoning with
  Certifying Proof Environments}

\author{Amira Henaien
\institute{LITA, Université de Lorraine, \\Ile du Saulcy, 57000, Metz, France}
\institute{Higher School of Communication of Tunis (Sup'Com), University of Carthage, Tunisia}
\email{amira.henaien5@etu.univ-lorraine.fr}
\and
Sorin Stratulat
\institute{LITA, Université de Lorraine, \\Ile du Saulcy, 57000, Metz, France}
\email{\quad sorin.stratulat@univ-lorraine.fr}
}
\def\titlerunning{Performing Certified Implicit Induction Reasoning}
\def\authorrunning{A. Henaien \& S. Stratulat}

\begin{document}
\maketitle
\begin{abstract}
  Largely adopted by proof assistants, the conventional induction
  methods based on explicit induction schemas are non-reductive and
  local, at schema level. On the other hand, the implicit induction
  methods used by automated theorem provers allow for lazy and mutual
  induction reasoning. In this paper, we present a new tactic for the
  Coq proof assistant able to perform automatically implicit induction
  reasoning. By using an automatic black-box approach, conjectures
  intended to be manually proved by the certifying proof environment
  that integrates Coq are proved instead by the Spike implicit
  induction theorem prover. The resulting proofs are translated
  afterwards into certified Coq scripts.
\end{abstract}
 
\section{Introduction} 
\label{sec:motivation}

Unbounded data structures like naturals and lists are abundant in
today's first-order specifications. Issued from the general principles
of Noetherian induction and its counter-positive version, the 
\emph{Descente Infinie} induction, various proof techniques have been devised to
effectively reason on well-founded posets. Given such a poset, the
soundness of the reasoning is guaranteed by the underlying
well-founded ordering which forbids infinite strictly decreasing
sequences of elements from the poset.

In a first-order setting, one can distinguish two classes of
Noetherian induction methods. The first class, of conventional
induction methods, is based on explicit induction
schemas~\cite{Burstall:1969pj}. An induction schema can attach to a
formula, called \emph{induction conclusion}, a set of formulas defined
as \emph{induction hypotheses} (IHs) to be used exclusively by the
offspring of the induction conclusion in the proof derivation. The
induction orderings are defined locally, at schema level, and can
differ from one schema to another. Such methods are widely-spread
among proof assistants because they can be easily integrated into
sequent-based inference systems as a separate inference rule. On the
other hand, it is not lazy, so it may happen that the IHs be defined
(sometimes long) before their use or to not be exploited at all. The
main challenges to overcome are the definition of the right IHs
followed by the guidance of the proof development related to the
induction conclusions in order to properly use the defined
IHs. Moreover, since only the offspring of the induction conclusion
can use its attached IHs, the mutual induction with other formulas
from the proof cannot be done naturally. The second class, of
implicit induction techniques, allows for lazy and mutual induction
because any formula instance from the proof can be used as IH, as long
as it is smaller than (and sometimes equal to) the current
conjecture. They fit better for reasoning as a working
mathematician~\cite{Wirth:2004rb}. On the other hand, the proof
derivations are reductive, requiring that the ground instances of
newly derived conjectures in the proof be smaller than (and sometimes
equal to) some ground instances of the current conjecture. In order to
satisfy the ordering constraints, the induction ordering is global, at
proof level. These techniques are originating from the Knuth-Bendix
completion algorithm~\cite{knuth1970swp,Musser:1980aq} and are usually
integrated into automatic theorem provers.

\sloppy{Proofs of similar conjectures done with the Spike implicit induction
prover~\cite{Bouhoula:1995wl,Stratulat:2001wo,Barthe:2003hv} and with
proof assistants like Coq~\cite{coq} and PVS~\cite{Shankar:2001zr}
have been previously
compared~\cite{Rusinowitch:2003ak,Barthe:2003hv}. We have witnessed}
that many of the conjectures can be automatically proved by
Spike, and concluded that the number of user interactions can
dramatically reduce if the proof assistants integrate implicit
induction techniques. To achieve this goal, we present a new Coq tactic
that performs lazy and mutual induction reasoning directly from the
certifying proof environment provided by Coq.  In an automatic way,
subgoals from a Coq script are exported to be proved by Spike. The
resulting Spike proofs are translated back into certified Coq scripts,
as shown by previous works~\cite{Stratulat:2010vn,Stratulat:2011kx}.

\paragraph{Related works} In the past, a lot of effort has been put
into adding implicit induction features to explicit induction proofs.
In \cite{Boyer:1988ve,Kapur:1996qw,Boulton:2000zr}, explicit
induction-based proof techniques systems have been extended to deal
with certain classes of mutually defined recursive functions.  Also,
Protzen~\cite{Protzen:1994ci} has defined a proof strategy to perform
lazy induction on particular explicit induction proofs. In the other
direction, Courant~\cite{Courant:1996gd} identified a class of
implicit induction proofs that can be reconstructed into conventional
induction proofs.  More recently, Voicu and Li~\cite{Voicu:2009vn}
proposed a \emph{Descente Infinie} tactic in Coq that identifies
repeated subgoals in a Coq script by analyzing their (partial) proof
terms.  A recursive function definition and an associated explicit
induction schema are issued from the analysis process.  The
implementation of the tactic is rather limited, requiring (a
restricted form of) rewriting due to the reductive nature of the
\emph{Descente Infinie} induction-based proof methods.  A naïve
automation algorithm was presented for simple inductive proofs.

\paragraph{Structure of the paper} The paper has 5 sections and one
appendix. After the introductory section, Section~\ref{sec:background}
introduces the main induction principles adapted for first-order
reasoning and shows how the implicit induction technique is
implemented in Spike. Section~\ref{sec:coq} briefly presents the Coq
system and describes the automatic tactic.  The implementation details
of the tactic and some experimental results are given in
Section~\ref{sec:implementation}.  The last section concludes.  The
statistics about the experimental part and the workflow of the
\texttt{Spike} tactic are given in the appendix.

\section{Induction principles and the Spike theorem prover}  
\label{sec:background}

\paragraph{Induction principles}
The induction principles of interest are instances of the general
Noetherian principle adapted for first-order logic. In
~\cite{Stratulat:2012uq}, they have been qualitatively distinguished
according to the kind of elements we are reasoning on, which can be
either terms or first-order formulas. 

The\emph{ term-based induction} principle considers that, in order to
prove a property $\phi$ over a set of (vectors of) terms $\cal E$, it
is enough to prove it for each element, knowing that we have the right
to assume as IH the fact that $\phi$ is true for any smaller element. The
explicit induction proof techniques, like the structural
induction~\cite{Burstall:1969pj}, implement the term-based induction
principle, hence they can be applied to prove only one property, in
our case $\phi$.

\begin{definition}[term-based induction]
\label{def:term-induction}
 $(\forall$ term vector
$\overline{m}\in {\cal E}, (\forall$ term vector $\overline{k} \in
{\cal E}, \overline{k} < \overline{m} \Rightarrow \phi(\overline{k}))
\Rightarrow \phi(\overline{m})) \Rightarrow\forall$ term vector $\overline{p} \in {\cal E}, \phi(\overline{p})$.
\end{definition}

On the other hand, several properties can be proved simultaneously by
using the \emph{formula-based induction} principle if the elements are
instead first-order formulas from a set $\cal E'$. The implicit induction is an example of
formula-based induction. It allows to a formula to be used as IH in
the proof of another formula as long as it is smaller, hence it is
able to perform mutual induction lazily.

\begin{definition} [formula-based induction]
\label{def:formula-induction}
$(\forall$ formula
$\delta\in {\cal E'}, (\forall$ formula $\gamma \in {\cal E'}, \gamma <
\delta \Rightarrow \gamma)\Rightarrow 
\delta) \Rightarrow \forall$ formula
$\rho \in {\cal E'}$\negthinspace, $\rho$. 
\end{definition}

The soundness of the two inductive principles is ensured if the
induction ordering is
well-founded, i.e., in our case, no infinite strictly decreasing
elements of $\cal E$ and $\cal E'$, respectively, can be built.

\paragraph{The Spike theorem prover} 

Spike~\cite{Bouhoula:1995wl,Stratulat:2001wo,Barthe:2003hv} is a
first-order theorem prover that can prove properties about conditional
specifications consisting of a set of axioms represented by
(conditional) equalities. In the past, Spike has been used to
automatize the validation process of several non-trivial applications
as the JavaCard Platform~\cite{Barthe:2003hv} and a telecommunications
protocol~\cite{Rusinowitch:2003ak}.

Its inference system is able to build
implicit induction proofs of the properties given as initial
conjectures by using inference rules that replace a conjecture from
the current state of the proof with a (potentially empty) set of new
conjectures. A Spike proof of a non-empty set of conjectures $E^0$ is a finite sequence of
states of the form \ensuremath{( E^0, \emptyset ) \vdash ( E^1, H^1 ) \vdash
  \cdots\vdash (H^{n-1}, H^{n-1})\vdash (\emptyset, H^n)}, where
\ensuremath{E^i ( i \in [0..n-1] )} are multisets of conjectures and
\ensuremath{H^i ( i \in [1..n] )} are multisets of previously treated
conjectures. 

The formula-based induction principle implemented by Spike considers
$\cal E'$ as the set of all ground instances of the conjectures
encountered in a proof but the IHs to be used during a proof step are
instances of formulas only from the current state. The induction
ordering is globally defined, at proof level. Moreover, the
inference rules are \emph{reductive} such that for each  ground
instance of the new conjectures generated in a proof step, there
should exist a smaller (and sometimes ordering-equivalent) logically
equivalent instance of a formula from the current state.

\section{Calling the Spike Prover from a Coq Script}  
\label{sec:coq}

\paragraph{The Coq proof assistant} Coq~\cite{coq} is a proof
assistant based on type theory. Its specifications can be written in
different ways. One of them is the functional style, using pattern
matching and recursion. In this case, the admissible functions have to
be well-typed and well-founded. The well-foundedness property of a
function can be checked by syntactical constraints on one of the
arguments of the function, called \textit{decreasing argument}. When
this argument is not explicitly stated, Coq tries to determinate it
but, in most of the cases, it is better to be given by the user. This
can be problematic when the functions are mutually recursively defined
and the user has to provide the induction ordering.

\paragraph{Example} In the following, we propose a Coq specification
of the mutually recursive functions \coqdocvar{even, oeven, odd} and
\coqdocvar{eodd} that take as arguments a natural and return a
boolean, using the Coq construction \coqdockw{Fixpoint} \ldots
\coqdockw{with}. The functions \coqdocvar{even} and \coqdocvar{oeven}
(resp. \coqdocvar{odd} and \coqdocvar{eodd}) are different
implementations of the predicate checking whether a natural is even
(resp. odd). The constructor symbols for the naturals are
\coqdocvar{0} and the successor $S$ and for booleans are
\coqdocvar{true} and \coqdocvar{false}. The constructor symbols are
free, i.e., there is no equality relation between them, and help to
define structural induction schemas from the function
definitions. They allow to clearly separate different branches in the
function definitions, using the \coqdockw{match}\ldots\coqdockw{with}
pattern matching construction.

\begin{multicols}{2}

\begin{coqdoccode}
\coqdocnoindent
\coqdockw{Fixpoint} \coqdocvar{even} (\coqdocvar{x}: \coqdocvar{nat}) :bool :=\coqdoceol
\coqdocindent{1.00em}
\coqdockw{match} \coqdocvar{x} \coqdockw{with}\coqdoceol
\coqdocindent{1.00em}
\ensuremath{|} 0 \ensuremath{\Rightarrow} \coqdocvar{true}\coqdoceol
\coqdocindent{1.00em}
\ensuremath{|} \coqdocvar{S} 0 \ensuremath{\Rightarrow} \coqdocvar{false}\coqdoceol
\coqdocindent{1.00em}
\ensuremath{|} \coqdocvar{S} (\coqdocvar{S} \coqdocvar{n}) \ensuremath{\Rightarrow} \coqdocvar{oeven} \coqdocvar{n}\coqdoceol
\coqdocnoindent
\coqdockw{end}\coqdoceol
\coqdocindent{1.00em}
\coqdockw{with} \coqdocvar{oeven} (\coqdocvar{x}: \coqdocvar{nat}) : \coqdocvar{bool} :=\coqdoceol
\coqdocindent{2.00em}
\coqdockw{match} \coqdocvar{x} \coqdockw{with}\coqdoceol
\coqdocindent{1.00em}
\ensuremath{|} 0 \ensuremath{\Rightarrow} \coqdocvar{true}\coqdoceol
\coqdocindent{1.00em}
\ensuremath{|} \coqdocvar{S} 0 \ensuremath{\Rightarrow} \coqdocvar{false}\coqdoceol
\coqdocindent{1.00em}
\ensuremath{|} \coqdocvar{S} (\coqdocvar{S} \coqdocvar{n}) \ensuremath{\Rightarrow} \coqdockw{if} \coqdocvar{odd} \coqdocvar{n} \coqdockw{then} \coqdocvar{false}\coqdoceol
\coqdocindent{7em}
\coqdockw{else} \coqdocvar{even} \coqdocvar{n}\coqdoceol
\coqdocnoindent
\coqdockw{end}\coqdoceol
\end{coqdoccode}

\begin{coqdoccode}
\coqdocindent{1.00em}
\coqdockw{with} \coqdocvar{odd} (\coqdocvar{x}: \coqdocvar{nat}) : \coqdocvar{bool} :=\coqdoceol
\coqdocindent{2.00em}
\coqdockw{match} \coqdocvar{x} \coqdockw{with}\coqdoceol
\coqdocindent{1.00em}
\ensuremath{|} 0 \ensuremath{\Rightarrow} \coqdocvar{false}\coqdoceol
\coqdocindent{1.00em}
\ensuremath{|} \coqdocvar{S} 0 \ensuremath{\Rightarrow} \coqdocvar{true}\coqdoceol
\coqdocindent{1.00em}
\ensuremath{|} \coqdocvar{S} (\coqdocvar{S} \coqdocvar{n}) \ensuremath{\Rightarrow} \coqdocvar{eodd} \coqdocvar{n}\coqdoceol
\coqdocnoindent
\coqdockw{end}\coqdoceol
\coqdocindent{1.00em}
\coqdockw{with} \coqdocvar{eodd} (\coqdocvar{x}: \coqdocvar{nat}) : \coqdocvar{bool} :=\coqdoceol
\coqdocindent{2.00em}
\coqdockw{match} \coqdocvar{x} \coqdockw{with}\coqdoceol
\coqdocindent{1.00em}
\ensuremath{|} 0 \ensuremath{\Rightarrow} \coqdocvar{false}\coqdoceol
\coqdocindent{1.00em}
\ensuremath{|} \coqdocvar{S} 0 \ensuremath{\Rightarrow} \coqdocvar{true}\coqdoceol
\coqdocindent{1.00em}
\ensuremath{|} \coqdocvar{S} (\coqdocvar{S} \coqdocvar{n}) \ensuremath{\Rightarrow} \coqdockw{if} \coqdocvar{even} \coqdocvar{n} \coqdockw{then} \coqdocvar{odd} \coqdocvar{n} \coqdoceol
\coqdocindent{7em}
\coqdockw{else} \coqdocvar{true}\coqdoceol
\coqdocnoindent
\coqdockw{end}.\coqdoceol
\end{coqdoccode}

\end{multicols}

Let us prove the theorem \\

\centerline{
\coqdockw{Theorem} \coqdocvar{even\_xx}: \ensuremath{\forall}
\coqdocvar{x}, \coqdocvar{even} (\coqdocvar{add x } \coqdocvar{x}) =
\coqdocvar{true}.
}

\centerline{}

\noindent where \coqdocvar{add} is the usual addition operator: 

\begin{multicols}{2}

\begin{coqdoccode}
\coqdocnoindent
\coqdockw{Fixpoint} \coqdocvar{add} (\coqdocvar{x} \coqdocvar{y}: \coqdocvar{nat}) :nat:=\coqdoceol
\coqdocnoindent
\coqdockw{match} \coqdocvar{x} \coqdockw{with}\coqdoceol
\coqdocnoindent
\ensuremath{|} 0 \ensuremath{\Rightarrow} \coqdocvar{y}
\end{coqdoccode}

\begin{coqdoccode}
\coqdoceol
\coqdocnoindent
\ensuremath{|} \coqdocvar{S} \coqdocvar{u} \ensuremath{\Rightarrow} \coqdocvar{S}(\coqdocvar{add} \coqdocvar{u} \coqdocvar{y})\coqdoceol
\coqdocnoindent
\coqdockw{end}.

\end{coqdoccode}
\end{multicols}

It can be shown that the trivial explicit induction schema which replaces
successively $x$ by $0$ and $S(y)$ does not work to successfully
finish the proof of the theorem. New user interaction is
required, for example, by adding other induction steps. The generation of useful
induction schemas  is not trivial for the general case. If the user
has no idea how the proof will be performed, the success of the proof
attempt may depend on: 

\begin{itemize}
\item the set of induction variables from the current conjecture,
\item the way the induction variables are instantiated. In our
  example, we have chosen 0 and S(n), but we could have had also 0,
  S(0) and S(S(n)), or 0, S(0), S(S(0)) and S(S(S(n))), etc. (in fact, there is an unbounded
  number of possibilities), and
\item the set of IHs attached to each induction conclusion and how
  they are applied further in the proof.  It may happen that additional IHs to be defined but not
used, or to miss crucial IHs in the definition of induction schemas.
\end{itemize}
A more detailed discussion about the `proof by induction' problems and
challenges can be found in~\cite{Gramlich:2005pf}.

\paragraph{The \texttt{Spike} tactic} An automatic solution that we propose
to the Coq users is to call instead the Spike theorem
prover. After providing two lemmas about some simple \texttt{add}
properties, the
proof can be completely done with the \texttt{Spike} tactic:

\centerline{}

\centerline{\texttt{Spike \textbf{[ordering constraints]}} }

\centerline{}

The tactic firstly generates a Spike specification from the analysis
of the Coq script, then Spike is executed to prove the theorem using
an induction ordering based on precedencies over the function symbols
given as arguments to the tactic. The precedencies can define
equivalence and strict ordering relations. In our case, the tactic and
its arguments are:

\centerline{}

\begin{center}
\begin{tabular}{ll}
\texttt{Spike} & \texttt{equiv}  [[\emph{even, oeven, odd, eodd}]] \\
& \texttt{greater} [[\emph{even, true, false, S, 0, add}], [\emph{add, S, 0}]].
\end{tabular}
\end{center}
\centerline{}

\noindent
It indicates to Spike that the symbols \coqdocvar{even, oeven, odd,
  eodd} are equivalent and that \coqdocvar{even}
(resp. \coqdocvar{add}) is greater than \coqdocvar{true, false, S, 0},
and \coqdocvar{add} (resp. \coqdocvar{S} and \coqdocvar{0}). The
implementation details of the \texttt{Spike} tactic are given in the next section.

\section{The Implementation of the \texttt{Spike} Tactic} 
\label{sec:implementation}

Tactics are at the heart of building proofs in Coq. They may use
elements of the current context of a given proof, as declarations,
definitions, axioms, hypotheses, lemmas and already proved theorems to
solve the current goal.  Coq allows the integration of new tactics by
two means: either by Ltac~\cite{Delahaye:2000vn}, a tactical language
for Coq, or by the OCAML programming language~\cite{ocaml}. The first
solution is less automated because of the difficulty to directly call
an external program, in occurrence the Spike prover, and of file
reading and writing operations. The second solution is based on an
interface between Coq and OCAML which lets developers integrate their
own tactics written in the fully featured OCAML language. On the other
hand, no safety control mechanisms are provided. Hopefully, the proof
terms built from tactics can be type-checked by the Coq kernel at the end
of a proof script. In addition, the lack of documentation and the
`not sufficiently explained' Coq source code usually make the
implementation task difficult. The second solution will be detailed
for proving \emph{even\_xx}.

Firstly, we present the full Coq script representing the axiomatic
translation of the function-based Coq script from
Section~\ref{sec:coq} and the proof of  \emph{even\_xx}:

\begin{coqdoccode}
 \coqdoceol \coqdocnoindent
\coqdockw{Axiom} \coqdocvar{add1 : $\forall$ x,  add 0  x = x}.
\coqdoceol
\coqdocnoindent\coqdockw{Axiom} \coqdocvar{add2 : $\forall$ x y, add (S x)  y  = S( add x y)}.
\coqdoceol
\coqdoceol
\coqdocnoindent\coqdockw{Axiom} \coqdocvar{even1} : \coqdocvar{even} 0 = \coqdocvar{true}.\coqdoceol
\coqdocnoindent
\coqdockw{Axiom} \coqdocvar{even2} : \coqdocvar{even} (\coqdocvar{S} 0) = \coqdocvar{false}.\coqdoceol
\coqdocnoindent
\coqdockw{Axiom} \coqdocvar{even3} : \ensuremath{\forall} \coqdocvar{x}, \coqdocvar{even} (\coqdocvar{S} (\coqdocvar{S} \coqdocvar{x})) = \coqdocvar{oeven} \coqdocvar{x}.\coqdoceol
\coqdocemptyline
\coqdocnoindent
\coqdockw{Axiom} \coqdocvar{oeven1} : \coqdocvar{oeven} 0 = \coqdocvar{true}.\coqdoceol
\coqdocnoindent
\coqdockw{Axiom} \coqdocvar{oeven2} : \coqdocvar{oeven} (\coqdocvar{S} 0) = \coqdocvar{false}.\coqdoceol
\coqdocnoindent
\coqdockw{Axiom} \coqdocvar{oeven3} : \ensuremath{\forall} \coqdocvar{x}, \coqdocvar{odd} \coqdocvar{x} = \coqdocvar{true} \ensuremath{\rightarrow} \coqdocvar{oeven} (\coqdocvar{S} (\coqdocvar{S} \coqdocvar{x})) = \coqdocvar{false}.\coqdoceol
\coqdocnoindent
\coqdockw{Axiom} \coqdocvar{oeven4} : \ensuremath{\forall} \coqdocvar{x}, \coqdocvar{odd} \coqdocvar{x} = \coqdocvar{false} \ensuremath{\rightarrow} \coqdocvar{oeven} (\coqdocvar{S} (\coqdocvar{S} \coqdocvar{x})) = \coqdocvar{even} \coqdocvar{x}.\coqdoceol
\coqdocemptyline
\coqdocnoindent
\coqdockw{Axiom} \coqdocvar{odd1} : \coqdocvar{odd} 0 = \coqdocvar{false}.\coqdoceol
\coqdocnoindent
\coqdockw{Axiom} \coqdocvar{odd2} : \coqdocvar{odd} (\coqdocvar{S} 0) = \coqdocvar{true}.\coqdoceol
\coqdocnoindent
\coqdockw{Axiom} \coqdocvar{odd3} : \ensuremath{\forall}
\coqdocvar{x}, \coqdocvar{odd} (\coqdocvar{S} (\coqdocvar{S}
\coqdocvar{x})) = \coqdocvar{eodd x}.\coqdoceol
\coqdocemptyline
\coqdocnoindent
\coqdockw{Axiom} \coqdocvar{eodd1} : \coqdocvar{eodd} 0 = \coqdocvar{false}.\coqdoceol
\coqdocnoindent
\coqdockw{Axiom} \coqdocvar{eodd2} : \coqdocvar{eodd} (\coqdocvar{S} 0) = \coqdocvar{true}.\coqdoceol
\coqdocnoindent
\coqdockw{Axiom} \coqdocvar{eodd3} : \ensuremath{\forall} \coqdocvar{x}, \coqdocvar{even} \coqdocvar{x} = \coqdocvar{true} \ensuremath{\rightarrow} \coqdocvar{eodd} (\coqdocvar{S} (\coqdocvar{S} \coqdocvar{x})) =  \coqdocvar{odd} \coqdocvar{x}.\coqdoceol
\coqdocnoindent
\coqdockw{Axiom} \coqdocvar{eodd4} : \ensuremath{\forall} \coqdocvar{x}, \coqdocvar{even} \coqdocvar{x} = \coqdocvar{false} \ensuremath{\rightarrow} \coqdocvar{eodd} (\coqdocvar{S} (\coqdocvar{S} \coqdocvar{x})) = \coqdocvar{true}.\coqdoceol
\coqdocemptyline
\coqdoceol
\coqdocnoindent
\coqdockw{Theorem}  \coqdocvar{addx0: $\forall$ x , add x 0  = x}.
\coqdoceol
\coqdocnoindent\coqdockw{Spike equiv} [[\coqdocvar{even, oeven, odd, eodd}]] \coqdockw{greater} [[\coqdocvar{oeven, add, S,
0, true, false}], [\coqdocvar{add, S, 0}]].
\coqdocnoindent\coqdockw{Qed.}
\coqdoceol
\coqdoceol
\coqdocnoindent
\coqdockw{Theorem} \coqdocvar{addS1: $\forall$ x y, add x (S y)  = S (add x y).}
\coqdoceol
\coqdocnoindent
\coqdocnoindent\coqdockw{Spike equiv} [[\coqdocvar{even, oeven, odd, eodd}]] \coqdockw{greater} [[\coqdocvar{oeven, add, S,
0, true, false}], [\coqdocvar{add, S, 0}]].
\coqdocnoindent\coqdockw{Qed.}
\coqdoceol
\coqdoceol
\coqdocnoindent\coqdockw{Theorem} \coqdocvar{even\_xx: $\forall$ x, even(add x x) = true}.
\coqdoceol
\coqdocnoindent\coqdockw{Spike equiv} [[\coqdocvar{even, oeven, odd, eodd}]] \coqdockw{greater} [[\coqdocvar{oeven, add, S,
0, true, false}], [\coqdocvar{add, S, 0}]].
\coqdocnoindent\coqdockw{Qed.}

\end{coqdoccode}

\centerline{}

The soundness of the translation can be checked for each axiom,
represented as a new theorem which is further proved. For example, the
axiom \emph{even3} can be validated by the theorem:

\centerline{}

\coqdocnoindent 
\coqdockw{Theorem} \coqdocvar{even3\_soundness} : \ensuremath{\forall} \coqdocvar{x}, \coqdocvar{even} (\coqdocvar{S} (\coqdocvar{S} \coqdocvar{x})) = \coqdocvar{oeven} \coqdocvar{x}.\coqdoceol
\coqdocnoindent\coqdockw{intro} \coqdocvar{x}. \coqdockw{simpl}. \coqdockw{trivial}. 
\coqdocnoindent\coqdockw{Qed.}

\centerline{}

The \texttt{Spike} tactic comes into four variants: 
\begin{description}
\item[Spike] {- the ordering constraints are inferred by Spike directly
    from the
    specification.}
\item[{Spike equiv [ $ S_1, \ldots , S_n $]}]  - each $S_i$
  ($i~\in~[1..n]$) is a set of equivalent symbols.
\item[{Spike greater [ $ S_1, \ldots , S_n$]}] - each $S_i$
\sloppy{  ($i\in[1..n]$) is of the form $\{symb_1, symb_2,  \ldots,  symb_n\}$ such
  that $symb_1$ is greater than any
    of the symbols from the set $\{symb_2,  \ldots, symb_n\}$.}
\item [{Spike equiv [ $ S_1, \ldots, S_n$  ] greater [  $ S^{'}_1, \ldots,
  S^{'}_n$ ]}] is  the combination of the two previous cases.
\end{description}

The tactic starts by extracting the Spike specification and the
conjecture to be proved from the current section, then calls Spike to
generate an implicit induction proof of the conjecture. If successful,
a Coq script is generated from the Spike proof representing the Coq
proof of an equivalent theorem to the current one. The script is
certified by the Coq kernel, then the equivalent theorem is applied to
solve the original Coq conjecture. The proof is completed only after
the Coq kernel has successfully type-checked the whole proof, i.e.,
when the \texttt{Qed} command ending the proof is executed. We will detail each
step of the tactic for the proof of \emph{even\_xx}, schematised in
Fig.~\ref{fig:schema-spikeT} from the appendix.

\paragraph{Extracting the Spike specification} A Spike specification
consists of two parts.  The first one describes the sorts with their
constructors and the defined function symbols by means of conditional
equalities playing the role of axioms. The second part influences the
way the proof will be performed. It includes the definition of the
precedencies over the function symbols, the definition of the
inference rules and the proof strategy. It also includes the
definition of the conjectures to be proved. The \texttt{Spike} tactic
is able to extract all this information from the current Coq proof
environment by the means of the "\emph{spike}" OCAML module. Any
Coq specification that intends to use the \texttt{Spike} tactic has
to declare previously this module, as follows: 

\centerline{}
\coqdockw{Declare ML Module} "\coqdocvar{spike}".
\centerline{}

\centerline{} The Coq specification should define the functions
axiomatically. One solution is to extract the axioms directly from the
fixpoint definitions, then validate them as shown in the previous
section for the axiom \emph{even3}.  The tactic starts by identifying
the conjecture to be proved from the current goal, then translates all
axioms existing in the current context, together with the sorts,
function symbols and constructors. All definitions and conjectures are
adapted to be accepted syntactically in Spike. In order to ease the
translation from Spike proofs to Coq scripts process, the generated
Spike specifications may contain inline Coq scripts, for example 
declaring the signature of the function symbols using
\texttt{Parameter}. For our example, the Spike specification starts
with:

\begin{description}
\item[sorts:]{ bool nat  ; }
\item[constructors:]{ 
\hfill
\begin{description}
\item{true :  $ \rightarrow $ bool; }
\item{false : $ \rightarrow $ bool; }
\item{ 0 :  $ \rightarrow $  nat; }
\item{S\_ : nat  $ \rightarrow $ nat;}
\end{description} 
}
\item[defined functions: ]{
\hfill
\begin{description}
\item{even\_  : nat $ \rightarrow $ bool;~\hspace{1cm} \$ \texttt{Parameter} even :
    nat $ \rightarrow $ bool.}
\item{oeven\_  : nat $ \rightarrow $ bool;\hspace{1cm} \$ \texttt{Parameter} oeven : nat $ \rightarrow $ bool.}
\item{odd\_  : nat $ \rightarrow $ bool;\hspace{1.3cm} \$ \texttt{Parameter} odd : nat
    $ \rightarrow $ bool.}
\item{eodd\_  : nat $ \rightarrow $ bool;\hspace{1.2cm} \$ \texttt{Parameter} eodd : nat $ \rightarrow $ bool.}
\item{add\_\_  : nat    nat $ \rightarrow $ nat ;\hspace{.6cm}  \$ \texttt{Parameter}
    add : nat  $ \rightarrow $  nat $ \rightarrow $   nat.}
\end{description}
}
\end{description}

\paragraph{Translating the Spike proof into Coq script} The Spike
prover is automatically executed on the generated specification using
a mode that can produce Coq script from a proof, as shown
in~\cite{Stratulat:2010vn,Stratulat:2011kx}. In the following, we only
recall the key steps of the translation process. In order to reproduce
the implicit induction reasoning, a comparison weight \coqdocvar{$W$}
is associated for each formula \coqdocvar{$F$}. When $F$ is
instantiated, $W$ has to be instantiated in the same way. This
is achieved by factorizing variables using functionals of the form
(\coqdockw{fun} \coqdocvar{$\overline{x} \Rightarrow $
}\coqdocvar{(F,W)}), where $\overline{x}$ is a vector of variables
shared between $F$ and $W$. For instance, the functional associated to
the conjecture labelled by $91$ in the Spike proof and being the
equivalent
of our theorem \emph{even\_xx} is: \\
\begin{coqdoccode}
\coqdoceol
\coqdockw{Definition} \coqdocvar{type\_LF\_91} :=  nat
\ensuremath{\rightarrow} (Prop * (List.list term)). \coqdoceol

\sloppy{\coqdockw{Definition F\_91} : \coqdocvar{type\_LF\_91}:= (fun  u1 \ensuremath{\Rightarrow} ((even (add u1 u1)) = true, (Term id\_even ((Term id\_add ((model\_nat u1):: (model\_nat u1)::nil))::nil))::(Term id\_true nil)::nil)).} \coqdoceol

\end{coqdoccode}

\centerline{}

The induction orderings used by Spike are syntactic and exploit the
tree representation of terms.  In Coq, we need to abstract the Coq
formulas into multisets of special terms built from a term algebra
provided by the COCCINELLE library~\cite{Contejean:2007kb}.  For
example, `id\_even' is the COCCINELLE equivalent of `even'. There are
also the \emph{model} functions that translate Coq terms into COCCINELLE
terms.  Finally, we get two main parts: firstly, the
specification part contains the current context with some
informations provided to COCCINELLE in order to create
operational term algebra. It includes other important steps, like  the description
of the induction ordering and its properties, as well as the
soundness proof of the underlying Noetherian induction principle.  
Secondly, the direct one-to-one translation of the
Spike proof for which the application of every inference rule has a corresponding Coq
script. 

An important contribution with respect to the previous works is the
replacement of the fixpoint-based functions by axioms. This
improvement allowed a more effective one-to-one translation of many
Spike rule applications, in particular for those dealing with
conditional equalities.  In the previous works, the method to identify
the conditional axiom used by a Spike rule consists in unfolding the
fixpoint definition of the function symbol defined by the conditional
axiom, then choosing the branch of the fixpoint definition that
validates the conditions of the axiom in the current proof
context. The generation of Coq script for the validation part is not
fully automatisable, so it may lead to the failure of the
\texttt{Spike} tactic. Moreover, thanks to the axiomatic
representation, many of the unconditional axioms can now participate
to automatize even more the translation process. Different rewriting
operations have been simplified:
\begin{
    enumerate}[i)]
\item{unconditional rewriting of the current goal: }{any unconditional
    axiom can be added to a rewriting base. For our example, the
    corresponding script is:

\centerline{}
\begin{coqdoccode} 
\coqdockw{Hint} \coqdocvar{Rewrite} \coqdocvar{addS1}
\coqdocvar{addx0} \coqdocvar{add1} \coqdocvar{add2}
\coqdocvar{even1} \coqdocvar{even2} \coqdocvar{even3}
\coqdocvar{oeven1} \coqdocvar{oeven2} \coqdocvar{odd1}
\coqdocvar{odd2} \coqdocvar{odd3} \coqdocvar{eodd1}
\coqdocvar{eodd2}: \coqdocvar{rewrite\_axioms}.\coqdoceol
\coqdockw{Ltac} \coqdocvar{rewrite\_ax}:= \coqdocvar{autorewrite} \coqdockw{with}  \coqdocvar{rewrite\_axioms}.\coqdoceol
\end{coqdoccode}

\centerline{} 

During the translation process, any application of unconditional
rewriting rule in the implicit proof will be replaced by $rewrite\_ax$.}
\item{conditional rewriting of the current goal: }{\sloppy{ it is
      translated into \\ \coqdocvar{rewrite}
    \coqdocvar{name\_of\_conditional\_axiom}.} }
\item{unconditional rewriting of a condition H of the current goal:}{
    it is translated into \coqdocvar{normalize}
    \coqdockw{with} \coqdocvar{rewrite\_axioms} \coqdockw{in}
    \coqdocvar{H}, where \coqdocvar{normalize} is a new Coq tactic we
    defined in order to rewrite H with the rewriting base
    \coqdocvar{rewrite\_axioms}.}
\end{enumerate} 

The theorem equivalent to \emph{even\_xx} is: \\

\centerline{
\coqdockw{Theorem} \coqdocvar{true\_91: $\forall$ u1, (even (add u1 u1)) = true.},}

\centerline{}

\noindent excepting that it is expected to be proved in the context of
the function symbols given as parameter to the Coq section including
\coqdocvar{true\_91}. Once the Coq script has been successfully
checked, it is imported into the current environment as a
library. Besides that, if a theorem needs some other theorems to be
proved in terms of lemmas, they should be put in the same section. The
\texttt{Spike} strategy will import the Coq script, apply the theorem
\coqdocvar{true\_91}, then discharge the parameters with function
symbols from the original Coq specification. The corresponding Coq
script is: 

\centerline{}
\coqdockw{Require Import} \coqdocvar{``Coq script
  with the proof of true\_91''}. 

\coqdockw{apply} \coqdocvar{true\_91}; 

\coqdockw{repeat} (\coqdockw{apply} \coqdocvar{even}
  $\mid$ \coqdockw{apply} \coqdocvar{oeven} $\mid$ \coqdockw{apply} \coqdocvar{odd} $\mid$ \coqdockw{apply} \coqdocvar{eodd} $\mid$ \coqdockw{apply}
  \coqdocvar{add}).

\paragraph{Extensions} The \texttt{Spike} tactic can only define
ordering constraints for the current proof. In order to deal with more
complex proofs,  two extensions are proposed:

\begin{itemize}
\item \sloppy{\texttt{SpikeWithFullind\_aug \textbf{[ordering constraints]}}} to
  integrate the augment technique  into the proof strategy. Presented
  in~\cite{Boyer:1988hb} and previously implemented in
  Spike~\cite{Armando:2002yd}, it allows to increase the factual base
  from the context of the simplifying conjectures.

\item \sloppy{\texttt{SpikeWithIndPriorities \textbf{[induction
      strategy]}\textbf{[ordering constraints]}}} to decide which terms
  from a conjecture can be instantiated. The induction strategy given
  as argument establishes a priority among the function symbols that
  may occur at root positions of these terms.

\end{itemize}

\paragraph{Statistics} We have also used the \texttt{Spike} tactic
and its extensions for other examples treated in previous
works~\cite{Stratulat:2010vn,Stratulat:2011kx}.
Table~\ref{fig:table-result} illustrates the number of lemmas (i.e.,
previously proved conjectures), hypotheses (i.e., not yet proved
conjectures), the employed tactic and whether it has used parameters or not, as
well as the execution time for some successfully proved conjectures
involved in the validation of a telecommunications
protocol~\cite{Rusinowitch:2003ak}. \texttt{Aug} and \texttt{Ind}
denote the use of \texttt{SpikeWithFullind\_aug} and
\texttt{SpikeWithIndPriorities} tactics, respectively. The proofs have
been performed with a MacBook Pro featuring 4 GB of RAM and a
2-core Intel processor i5 at 2,5 GHz. The full Coq scripts can be
found on Spike's website
\url{http://code.google.com/p/spike-prover/}. \footnote{Other examples
  of Coq scripts using the \texttt{Spike} tactic can be accessed from
  the website.}

\section{Conclusions and Future Works}  
\label{sec:conclusions}

We have presented the \texttt{Spike} tactic and some of its extensions
to automatically perform implicit induction by the Coq proof
assistant. Using a black-box approach, Coq is now able to certify
non-trivial conjectures whose proofs may require multiple induction
steps, as well as lazy and mutual induction reasoning. As a case
study, conjectures involved in the validation of a non-trivial
application have been successfully and directly certified by Coq using
the \texttt{Spike} tactic. The proofs of more than 60\% of them have
been performed completely automatically, i.e., the Coq user does not
need to provide any argument to the tactic. On the other hand, its
application is limited to Coq specifications transformable
into conditional specifications whose axioms can be oriented into
rewrite rules.

In the near future, we intend to automatize the translation process of
a fixpoint-based function definition into axioms, as well as to define a set of Coq tactics that
can build implicit inductive proofs directly into Coq either automatically, for example by
simulating Spike's behaviour, or interactively. For the last case,
a proof environment has to be designed to facilitate the user's access
to crucial information from the whole proof, in particular from the proof branches
other than that corresponding to the current goal. The idea is to use
this information to discover lazily the IHs allowing to perform  mutual
induction reasoning.   In a longer perspective, we will test whether these ideas can
be extrapolated to cyclic proofs and applied to the
recent induction methods proposed in~\cite{Stratulat:2012uq}.


\appendix
\section{Tables and Figures}

\begin{table}[h]
\small 
\begin{center}
\begin{tabular}{rlccccc}
  \hline
  \texttt{\#\mbox{\,\,}}& \texttt{Name} & \texttt{Lemmas}
  &\texttt{Hypotheses} & \texttt{Tactic}& \texttt{Parameters} &
  \texttt{~~~Time (s) }  
  \\ \hline 

1.& firstat\_timeat & 1 &0& \texttt{Spike} & no & 3.021\\
2. &firstat\_progat &1 &0& \texttt{Spike} & no & 3.159\\
3.&sorted\_sorted &0&0&\texttt{Spike}& no & 2.295 \\
4.&sorted\_insat1 &4&0&\texttt{Aug}& yes &14.233\\
5.&sorted\_insin2 &4&0&\texttt{Aug}& yes &15.400\\
6.& sorted\_e\_two &0&0&\texttt{Spike}& no&2.039\\
7.& member\_t\_insin &2&0&\texttt{Spike}& yes&8.558\\
8.&membert\_insat &3&0&\texttt{Spike}& no&11.317\\
9. &member\_firstat &2&0&\texttt{Spike}& no& 8.471\\
10. &timel\_insat &0&0&\texttt{Spike}& no&2.402\\
11. &erl\_insin &0&0&\texttt{Spike}& yes&2.512\\
12. &erl\_insat &0&0&\texttt{Spike}& no&2.488\\
13. &erl\_prog &2&0&\texttt{Spike}& yes&11.768\\
14. &time\_progat\_er &1&0&\texttt{Spike}& no&4.238\\
15. &timeat\_tcrt &0&0&\texttt{Spike}& no&3.225\\
16. &timel\_timeat\_max &2&1&\texttt{Aug} & yes&9.191\\
17. &null\_listat &1&0&\texttt{Spike}& no& 4.143\\
18. &null\_listat1 &0&0&\texttt{Spike}& no&1.958\\
19. &cons\_insat &0&0&\texttt{Spike}& no&1.987\\
20. &cons\_listat &0&0&\texttt{Spike}& no& 2.004\\
21. &progat\_timel\_erl &3&0&\texttt{Aug} & yes&11.621\\
22. &progat\_insat &3&1&\texttt{Aug}& yes&44.422\\
23. &progat\_insat1 &3&0&\texttt{Aug}& yes&17.055\\
24. &timel\_listupto &0&0&\texttt{Spike}& no&2.295\\
25. &sorted\_listupto  &3&0&\texttt{Aug}& yes&13.042\\
26. &time\_listat &1&0&\texttt{Spike}& no&5.720\\ 
27. &sorted\_cons\_listat &3&1&\texttt{Ind}& yes&16.699\\
28. &null\_wind2 &0&1&\texttt{Spike}& no&3.382\\
29. &timel\_insin1 &1&0&\texttt{Spike}& yes&4.456\\ 
30. &null\_listupto1 &0&0&\texttt{Spike}& no&1.949\\ 
31. &erl\_cons &0&0&\texttt{Spike}& no&2.415\\
32. &no\_time &1&1&\texttt{Spike}& no&7.129\\
33. &final &2&1& \texttt{Spike}& no & 8.330

  \end{tabular}
\end{center}
\centerline{}
\caption{\label{fig:table-result} Statistics about the ABR proofs.}
\end{table}

\begin{landscape}
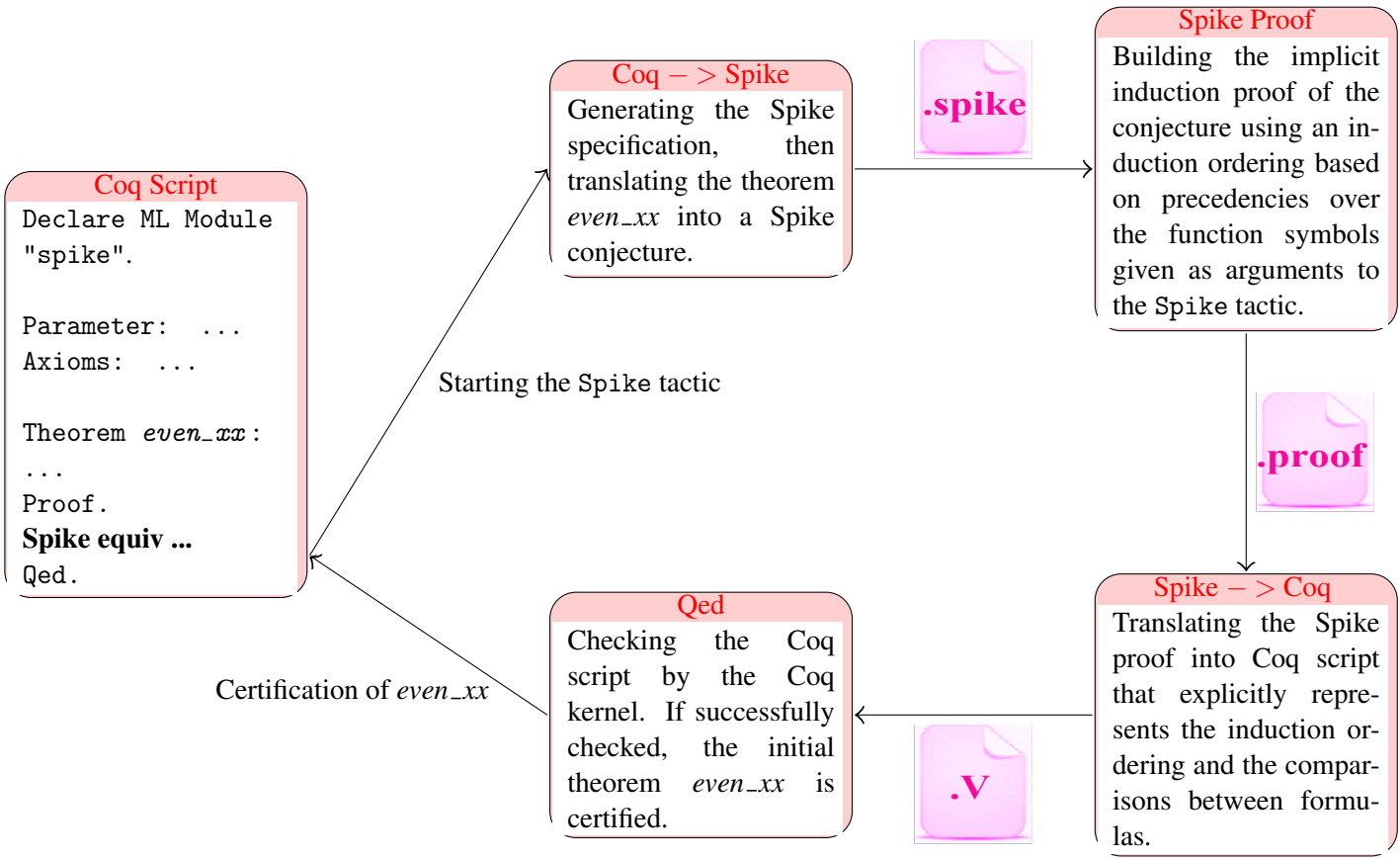
\begin{figure}
\begin{tikzpicture}[node distance = .2\textwidth]

\nouvelleN{}{A} {\centerline{Coq Script}
\colorbox{white}{
\begin{minipage}{.9\textwidth}
\normalsize {
\begin{coqdoccode}
\coqdockw{Declare ML Module "spike"}.\coqdoceol

\coqdoceol

\coqdockw{Parameter: \ldots}\coqdoceol
\coqdockw{Axioms: \ldots}\coqdoceol \coqdoceol
\coqdockw{Theorem  \coqdocvar{even\_xx}: \ldots}\coqdoceol
\end{coqdoccode}
\texttt{Proof.}\coqdoceol
\textbf{Spike equiv ...}\coqdoceol 
\texttt{Qed.}\coqdoceol 
}\end{minipage}}
};

\nouvelleN{right= of A.north east}{B}{
\centerline{Coq $->$ Spike}
\colorbox{white}{
\begin{minipage}{.9\textwidth}
\normalsize {
Generating the Spike specification, then translating the theorem \emph{even\_xx} into a Spike conjecture.}
\end{minipage}}
};
\draw[
    decoration={markings,mark=at position 1 with {\arrow[scale=2]{>}}},
    postaction={decorate},
    shorten >=0.4pt
    ] [->] (A.-48) to node [auto,swap,xshift=0mm] {Starting the \texttt{Spike} tactic} (B.west);

\nouvelleN{right= of B} {C} {\centerline{Spike Proof}
\colorbox{white}{
\begin{minipage}{.9\textwidth}
\normalsize {
Building the implicit induction proof of the conjecture using an induction ordering based on precedencies over the function symbols given as arguments to the \texttt{Spike} tactic.
}
\end{minipage}}
};

\draw[
    decoration={markings,mark=at position 1 with {\arrow[scale=2]{>}}},
    postaction={decorate},
    shorten >=0.4pt
    ] [->] (B.east) to node [auto] {\includegraphics[width=.1\textwidth,height=.1\textheight]{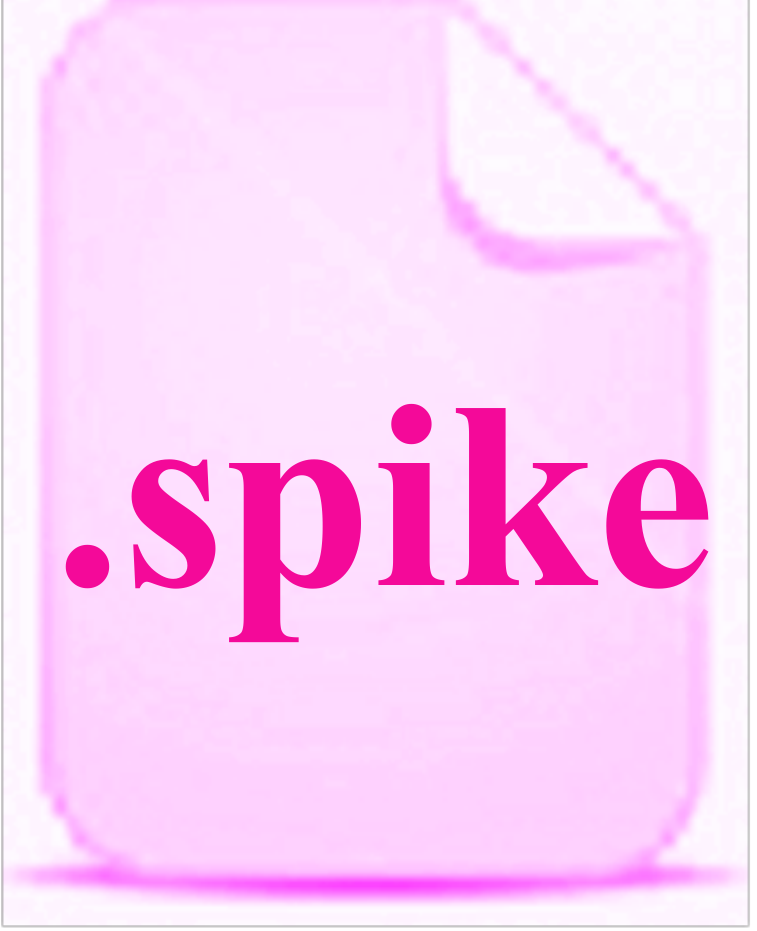}} (C.west);

\nouvelleN{below= of C} {D} {
\centerline{Spike $->$ Coq }
\colorbox{white}{
\begin{minipage}{.9\textwidth}
\normalsize {
Translating the Spike proof into Coq script that explicitly represents the induction ordering and the comparisons between formulas.
}
\end{minipage}}
};

\draw[
    decoration={markings,mark=at position 1 with {\arrow[scale=2]{>}}},
    postaction={decorate},
    shorten >=0.4pt
    ] [->] (C.south) to node [auto] {\includegraphics[width=.1\textwidth,height=.1\textheight]{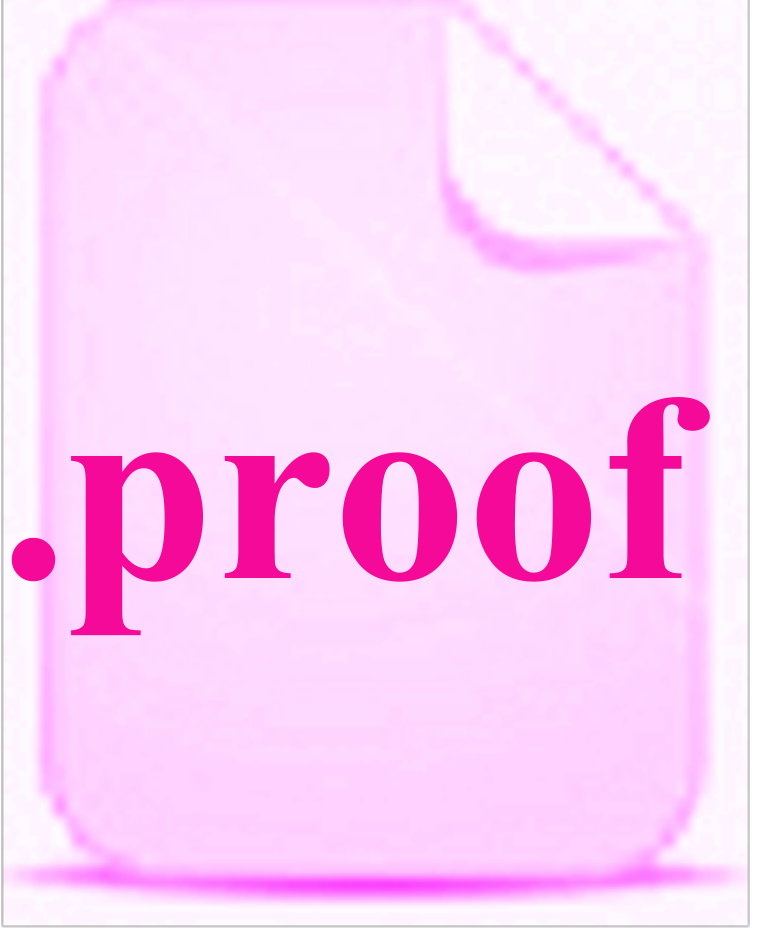}} (D.north);

\nouvelleN{
  left= of D
} {E} {\centerline{Qed}
\colorbox{white}{
\begin{minipage}{.9\textwidth}
\normalsize{Checking the Coq script by the Coq kernel. If successfully checked, the initial theorem \coqdocvar{even\_xx} is certified.}
\end{minipage}}
};

\draw[
    decoration={markings,mark=at position 1 with {\arrow[scale=2]{>}}},
    postaction={decorate},
    shorten >=0.4pt
    ] [->] (D.west) to node [auto] {\includegraphics[width=.1\textwidth,height=.1\textheight]{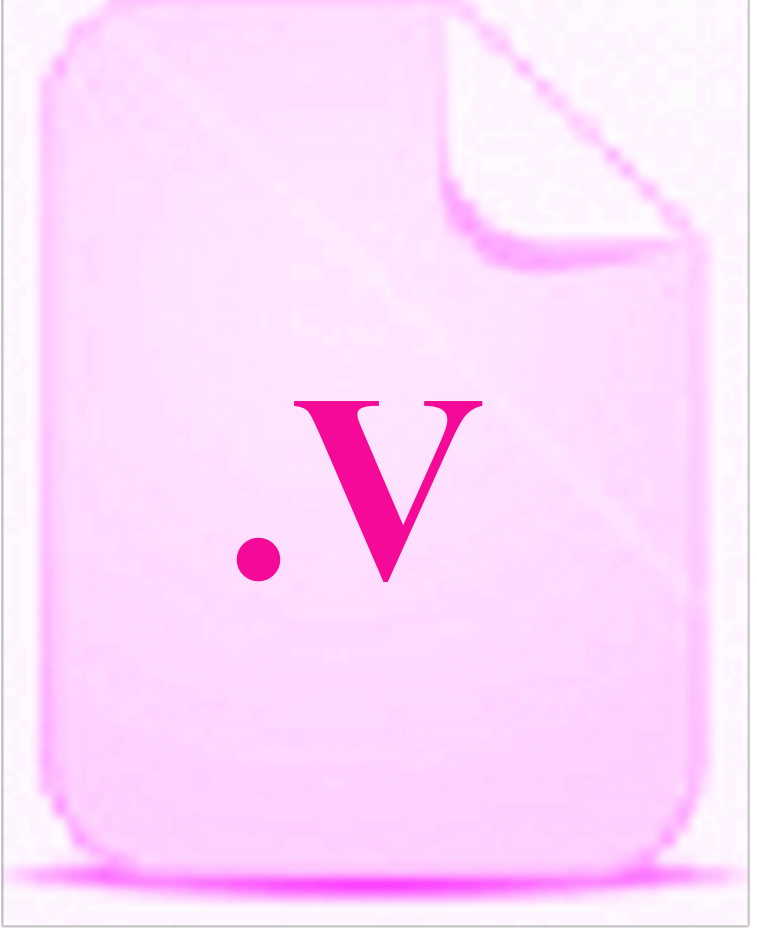}} (E.east);

\draw[
    decoration={markings,mark=at position 1 with {\arrow[scale=2]{>}}},
    postaction={decorate},
    shorten >=0.4pt
    ] [->] (E.west) to node [auto,swap,xshift=-30mm,yshift=-10mm] {Certification of \coqdocvar{even\_xx} } (A.-48);

\end{tikzpicture}
 \caption{\label{fig:schema-spikeT}The integration workflow of the \coqdocvar{even\_xx} Spike proof into Coq using the
   \texttt{Spike} tactic.}
\end{figure}
\end{landscape}

\end{document}